# Correlation between Exciton Dynamics and Spin Structure in van der Waals Antiferromagnet $NiPS_3$

Kang Wang[1]#, Yingchen Peng[1]#, Boying Huang[2]#, Chun Zhou[3], Qianlu Sun[4], Fujie Tang*[2], Zhenglu Li[5], Weigao Xu[4], Kezhao Du[6], Xingzhi Wang[3], Ye Yang*[1]

1. The State Key Laboratory of Physical Chemistry of Solid Surfaces, iChEM (Collaborative Innovation Center of Chemistry for Energy Materials), College of Chemistry & Chemical Engineering, Xiamen University, Xiamen 361005, China
2. Pen-Tung Sah Institute of Micro-Nano Science and Technology, Xiamen University, Xiamen 361005, Fujian Province, China
3. Department of Physics, Xiamen University, Xiamen, 361005, China
4. Key Laboratory of Mesoscopic Chemistry, School of Chemistry and Chemical Engineering, Nanjing University, Nanjing 210023, China
5. Mork Family Department of Chemical Engineering and Materials Science, University of Southern California, Los Angeles, CA 90089, USA
6. Fujian Provincial Key Laboratory of Advanced Materials Oriented Chemical Engineering, Fujian Normal University, Fuzhou 350007, P. R. China

# These author contributed equally in this work.
*Corresponding author.
Email: tangfujie@xmu.edu.cn
Email: ye.yang@xmu.edu.cn

**Abstract:** The emerging magnetic van der Waals (vdW) materials provide a platform for exploring novel physics regarding magnetism in low dimensions and developing ultrathin spintronic applications. Here, we investigate the ultrafast dynamics of excitons in a vdW $NiPS_3$ crystal. The temporal evolution of the transient reflection spectra indicates that the spin-correlated exciton is formed through photocarrier localization, the rate of which is independent of the magnetic degrees of freedom. However, the recombination rate of these excitons is connected with the long-range magnetic order, and this connection probably arise from a spin-flip rooted in the underlying antiferromagnetic background during the recombination. Our findings uncover intertwined coupling between carrier, lattice and spin degrees of freedom in $NiPS_3$, which may pave the path toward ultrafast optical manipulation of spin-related quantum states in vdW antiferromagnets.

The layered van der Waals (vdW) antiferromagnets have attracted intense research interest owing to their promising application in ultrathin spintronics and magnonics.[1-5] These emerging materials also offer an ideal platform for studies of magnetism in two dimensions.[6-14] Recently, several intriguing physical phenomena closely connected with magnetic ordering have been observed in these materials. For instance, excitons coupled with the magnetic degrees of freedom can be optically generated in nickel phosphorous trisulfide ($NiPS_3$) crystals, giving rise to ultranarrow and linearly-polarized emission and absorption.[4,14-18] The coupling between magnetic degrees of freedom and various electronic states [19,20] or phonons[9,12,15,21-26] in vdW antiferromagnets has also been reported, and these findings provide potential methods to optically manipulate the physical properties associated with the spin structures.[2,3,8,11,16,20,27-29] The correlation between excited states and magnetic ordering in the vdW antiferromagnets is at the core of these intriguing phenomena, insights of which are critical to finding ultrafast spin manipulation pathways.

Incompletely filled *d*-orbitals from the transition metal ions incorporated in the vdW antiferromagnets lead to complex nature of the optical excitation, and unveiling the correlation between the excitons with the magnetic order is challenging. Several different mechanisms have been proposed to explain the spin-correlated excitons in the vdW $NiPS_3$ antiferromagnet. These excitons have been recognized as an excitation from a Zhang-Rice triplet ground state to the singlet excited state.[4,19,30-32] However, a recent study suggested that the spin-correlated exciton could be theoretically constructed by considering the Hund exchange and inter-cluster hopping, invoking that the Zhang-Rice model was not necessary.[33] The resonant inelastic X-ray scattering experiment also uncovered that the excitons in antiferromagnetic $NiPS_3$ comprised more correlated *d*-orbitals than Zhang-Rice-like (p-d hybridized) orbitals, and thus the Hund's exchange interaction might primarily account for the exciton formation.[14,34] An alternative formation mechanism was also offered by a computational study that suggested a bound state at local lattice defects exhibiting spectral features similar to the spin-correlated excitons.[35] The localized nature can be derived from all of the above physical pictures: the Zhang-Rice states are formed by a *d*-orbital hole of a nickel ion and a hole in the *p*-shell of the surrounding sulfur ligands, and thus the excitonic wave function is expected to be strongly confined around a $NiS_6$ octahedron;[36] the strong Hund's coupling in the antiferromagnetic background can also lead to a highly localized exciton.[37,38] Moreover, considering that the small lattice polaron formation is ubiquitous in a Mott or charge-transfer insulator with *d*-orbital bands, the lattice polarization may also induce exciton localization.[39,40] In addition to a lattice polaron, the exciton could be dressed by magnons to form a magnetic polaron.[33] The coupling between the excitons and local lattice or spin structure may potentially complicate the exciton dynamics in the vdW magnets.

Here, we investigated the ultrafast exciton dynamics and its correlation with magnetic orders in a vdW $NiPS_3$ antiferromagnet primarily using transient reflection (TR) spectroscopy. The temporal evolution of TR spectra indicates that the spin-correlated excitons result from photocarriers localization occurring on a sub-ps time scale. The exciton formation rate is independent of the magnetic order, suggesting that the delocalized photocarriers and the spin structure are nearly uncoupled. In contrast, the temperature dependence of the exciton recombination rate manifests a close connection between the recombination dynamics and the magnetic order. Our findings offer unique insights into the intertwined coupling between carrier, lattice and spin degrees of freedom in vdW antiferromagnets.

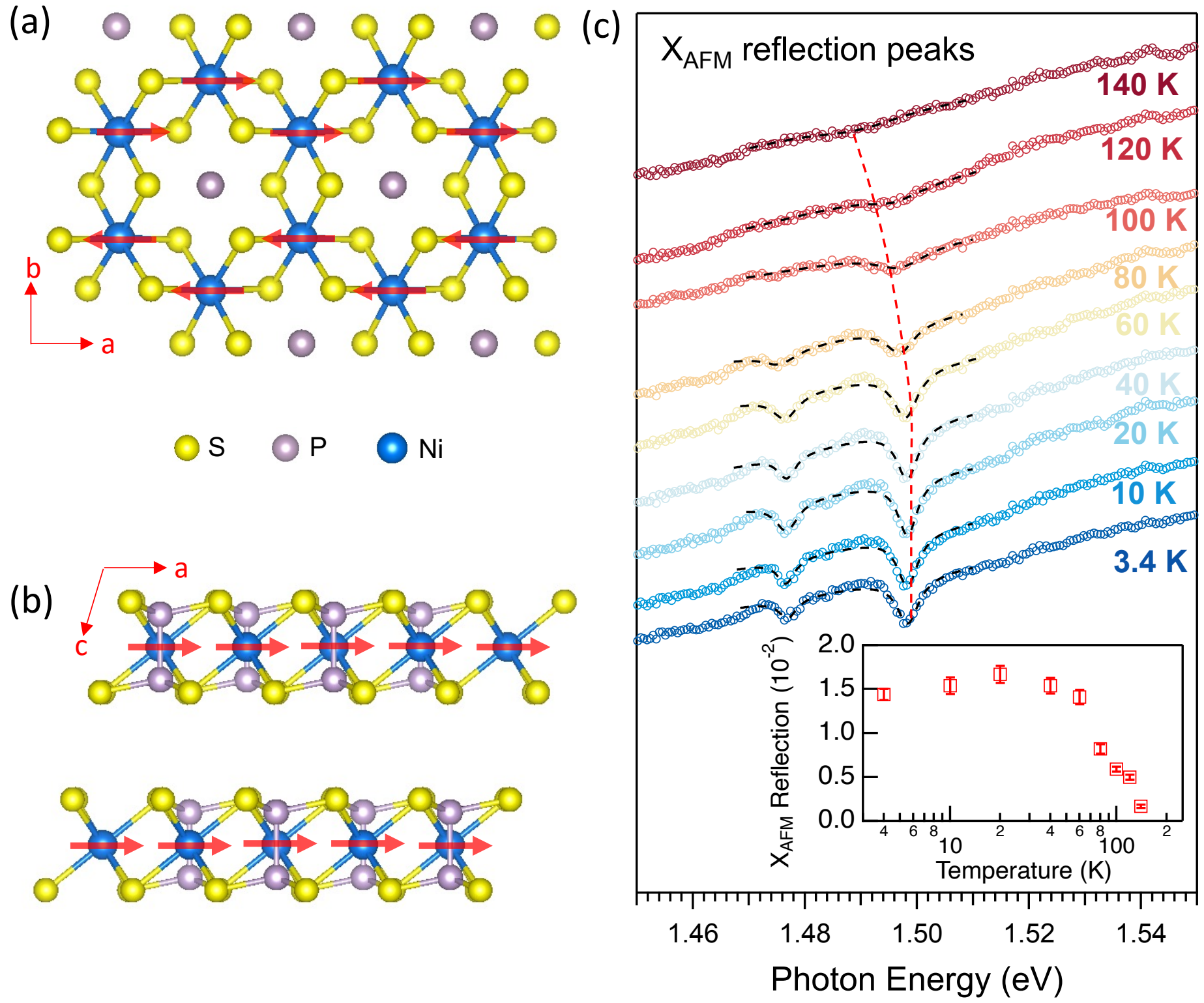

**Figure 1. Crystal structure and reflection spectra of $NiPS_3$.** (a) Illustration of the zig-zag ferromagnetic order along *a*-axis. (b) Illustration of the interlayer spin configuration. Red arrows represent the spins. (c) Steady-state reflection spectra measured at indicated temperatures. These traces are vertically offset for a better comparison. The $X_{AFM}$ reflection resonances with the magnon sideband (marked by red-dash curve) are fit by the Lorentzian function (black-dash curves). Inset shows the amplitude of the magnon sideband given by the fitting versus temperature, and the error bars represent the fitting uncertainty.

**Optical transitions of excitons coupled with the antiferromagnetic order ($X_{AFM}$).** A high-quality $NiPS_3$ crystal was obtained using a chemical vapor transport growth method[41] and subsequently cleaved to achieve a smooth surface for spectroscopic measurements. The X-ray diffraction pattern (Fig. S1) suggests that the exposed surface belongs to the (001) lattice plane containing *a*-axis and *b*-axis. As illustrated in Fig. 1a, nickel ions, octahedrally coordinated by sulfur ions, form a honeycomb lattice in the *ab* plane. When the temperature drops below $T_N$ (~ 155K),[42,43] spins of the Ni ions along the *a*-axis are aligned ferromagnetically in a zigzag chain, and these spin chains are arranged in an antiferromagnetic configuration along the *b*- axis.[19,42,43] Fig.1b shows the side view of the $NiPS_3$ layers, which are stacked ferromagnetically along the *c*-axis.

$NiPS_3$ is classified as a charge-transfer insulator with a bandgap energy ($E_g$) of ~1.8 eV.[19,44] The optical transitions above the bandgap possess a charge-transfer character, i.e., from sulfur *p*-orbitals to nickel *d*-orbitals, and the *d*-*d* transitions of local nickel ions account for the spectral features below $E_g$.[30,45,46] In the sub-bandgap region of the reflection spectra, two sharp peaks located at 1.47 eV and 1.49 eV tend to vanish when temperature approaches to $T_N$ (Fig. 1c), and such dependence (inset of Fig. 1c) indicates the intimate connection with the AFM order. The original spectra without offset are shown in Fig. S2. These reflection peaks align with the previously reported sharp absorption or emission peaks, which have been attributed to the AFM-order correlated exciton ($X_{AFM}$),[4,15-17,22,30,34,35,47] and the peaks at 1.47 eV and 1.49 eV have been identified as the $X_{AFM}$ resonance and magnon sideband, respectively.[4,22,28,35,36]

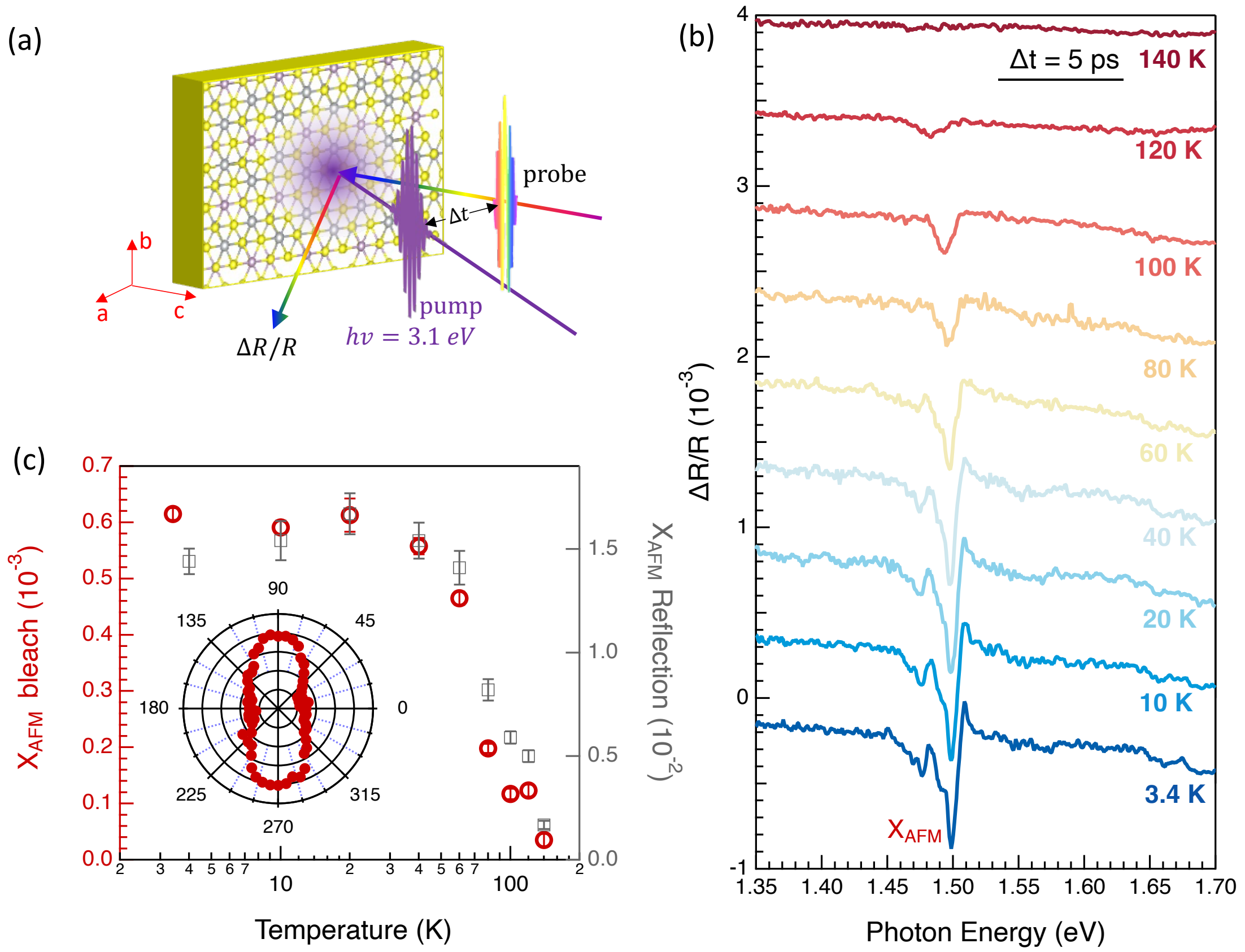

**Figure 2. Photoinduced $X_{AFM}$ TR feature associated with antiferromagnetism.** (a) Schematic illustration of TR spectroscopy. (b) TR spectra recorded at 5 ps for the indicated temperatures. The traces (except 3.4 K) are vertically offset for a better comparison. (c) The temperature dependent amplitude of the $X_{AFM}$ bleach recorded at the magnon sideband position (red circles). The peak amplitude was obtained by subtracting the broad background. The error bars represent the standard deviation obtained from three set repeated measurements. Also compared is the temperature-dependent steady-state $X_{AFM}$ reflection (magnon sideband, grey squares). Inset displays the probe-polarization dependence of the $X_{AFM}$ bleach amplitude measured at 3.4 K.

**Temperature dependence of photoinduced $X_{AFM}$ TR feature**. A direct approach to investigate the dynamics of $X_{AFM}$ excitons is to trace their time-resolved spectral features. As illustrated in Fig. 2a, the photon energy of the monochromatic pump (3.1 eV) was much greater than $E_g$, and the photoinduced reflection change ($\Delta R/R$) was detected by a supercontinuum probe with linear polarization. The TR spectra recorded at a representative delay (e.g., 5 ps) for different temperatures are displayed in Fig. 2b. When the temperature drops well below $T_N$, a pair of sharp peaks emerges at ~ 1.49 eV and ~1.47 eV, corresponding to the $X_{AFM}$ resonance with its magnon sideband shown in the steady-state reflection spectra. These TR features are attributed to the $X_{AFM}$ bleach because they are predominantly caused by the reduction of the $X_{AFM}$ oscillator strength (Fig. S3).[39,47] Like the steady-state $X_{AFM}$ resonances, the $X_{AFM}$ bleach progressively fade away when the temperature approaches to $T_N$. The temperature dependence of the $X_{AFM}$ bleach (red circles, Fig. 2c) resembles that of the $X_{AFM}$ steady-state reflection (grey squares, Fig. 2c). Moreover, the $X_{AFM}$ bleach also displays large anisotropy with respect to the probe polarization (inset, Fig. 2c), in line with anisotropy of the $X_{AFM}$ steady-state absorption or emission peaks that probably stem from the strong coupling between the exciton and anisotropic spin structure.[4,15,16,35] The probe-polarization dependent TR spectra measured at 3.4 K are displayed in Fig. S4.

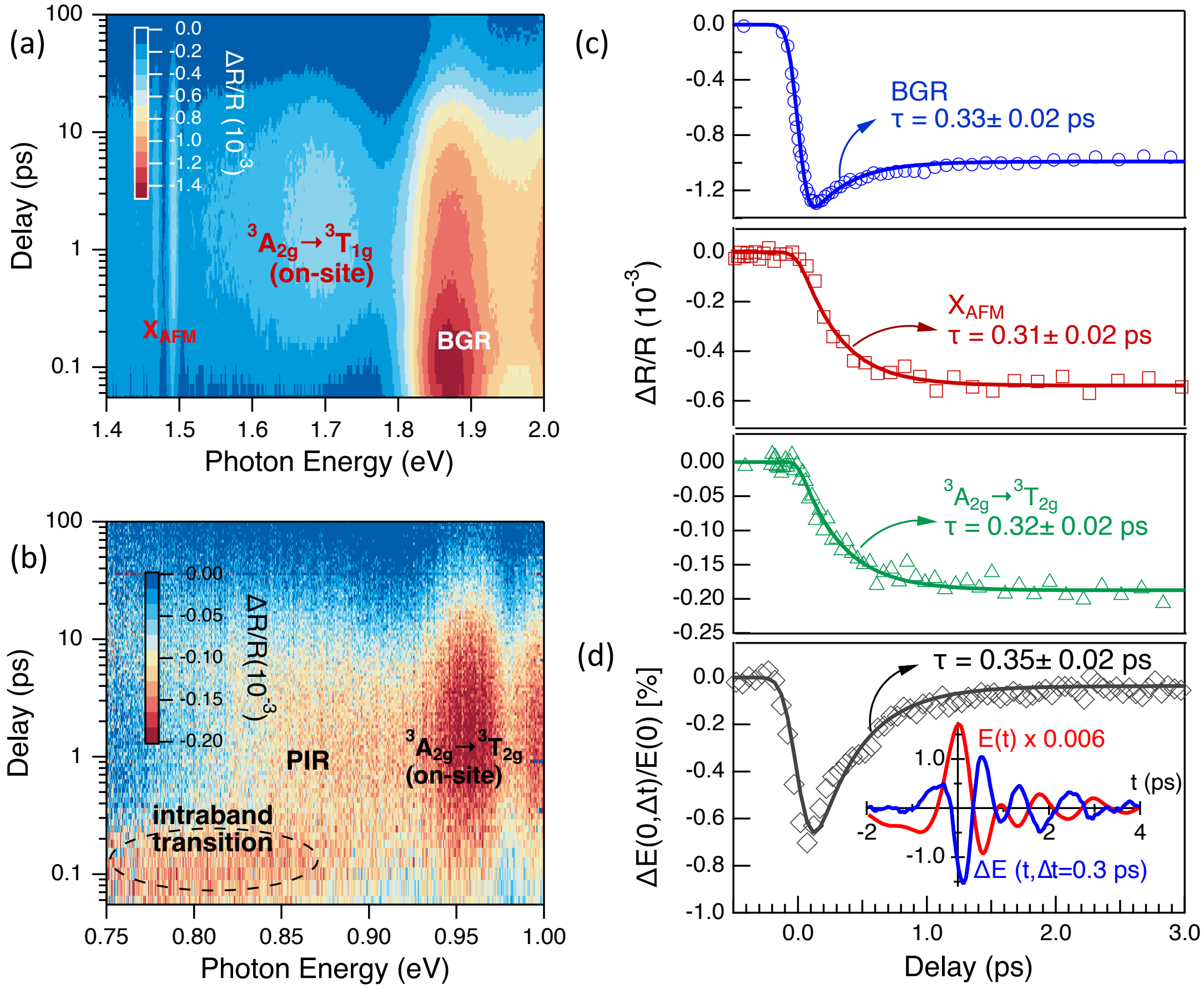

**Figure 3. Dynamics of the spin-correlated excitons.** Pseudo-color images of the transient reflection spectra at different delays probed in (a) visible and (b) near-infrared regions. The vertical and horizontal axes represent the delay time and probe photon energy, respectively. The signal magnitude is indicated by the color scale bar. (c) Comparison of the kinetics of the indicated spectral features at early delays. (d) Transient THz kinetics corresponds to the relative change in THz electric field in the time-domain. Inset shows the transient $[\Delta E(t;\Delta t)]$ and equilibrium $[E(t)]$ THz waveforms. The TR and THz experiments were conducted at 3.4 K. Symbols and solid curves represent experimental data and fitting curves, respectively.

**Localized nature of $X_{AFM}$ excitons.** As shown in Fig. 3a and 3b, the pseudo-color image of the TR spectra in the visible and near-infrared regions explicitly display the temporal evolution of the $X_{AFM}$ bleach along with other TR features. Beside the sharp $X_{AFM}$ bleach, a prominent TR feature near $E_g$, attributed to a spectral red-shift caused by the bandgap renormalization (BGR),[39,48] is formed instantaneously and then experiences a partial decay within 1 ps. The TR peak located at 1.7 eV corresponds to the spin-allowed ${}^3A_{2g} \rightarrow {}^3T_{1g}$ transition of local $Ni^{2+}$ ions, and another peak centered at 0.97 eV is assigned to the shoulder peak of the ${}^3A_{2g} \rightarrow {}^3T_{2g}$ transition due to lattice-distortion induced splitting of the ${}^3T_{2g}$ level. [14,27,45,46] A relatively broad photo-induced reflection (PIR) emerges below the shoulder peak, which likely arises from the optical transitions associated with excited states.[39] As shown latter, those on-site *d*-*d* transitions and PIR share common dynamic behaviors with the $X_{AFM}$ bleach. All these TR features are labeled in Fig. 3a and 3b.

Since the pump photon energy is greater than $E_g$, the optical excitation initially creates delocalized holes and electrons in the valence and conduction bands, respectively. The many-body interaction of these delocalized carriers narrows the bandgap, accounting for the short-lived component of BGR. Furthermore, the intraband transition of these photocarriers should be responsible for a broad TR feature beyond 0.85 eV that quickly vanishes in parallel with the partial decay of BRG.[49] The rapid decay of BGR and intraband transition represents quick depopulation of the delocalized photocarriers through localization or recombination. The recombination returns the system to the ground state, whereas the decay of these TR features is accompanied by the formation of the $X_{AFM}$ bleach and *d*-*d* transitions, instead of a complete spectral recovery. Hence, we can exclude the charge recombination and ascribe the fast decay to the photocarrier localization. The localized state is likely coupled with phonons, corroborated by a series of phonon replicas appearing below $E_g$ on the TR spectra (Fig. S5).[15,22] Prior to the localization, the photocarriers might be in the form of large polarons because of the strong electron-phonon coupling[15,22,35] or electron-magnon coupling[2,4,27,28].

To quantitively depict the dynamic interplay between these TR spectral features, their kinetics at early delays are displayed in Fig. 3c. The resonance and the magnon sideband of $X_{AFM}$ bleach share a common kinetics, which are also consistent with the formation of $^3A_{2g} \rightarrow {}^3T_{1g}$ (Fig. S6). The $X_{AFM}$ kinetics in Fig. 3c is recorded at the magnon sideband. The single exponential fitting of these kinetics indicates excellent consistence between the fast BGR decay and the growth of the $X_{AFM}$ bleach and $^3A_{2g} \rightarrow {}^3T_{2g}$. This coincidence suggests that the photocarriers collapse into localized spin-correlated excitons that bleach the $X_{AFM}$ transition and modify the on-site *d-d* transitions. The $X_{AFM}$ bleach exhibits a single exponential decay pattern that is nearly independent of the pump fluence (Fig. S7), characteristic of a single-particle recombination mechanism and thus consistent with the exciton recombination picture. The decay kinetics of $X_{AFM}$ bleach and $^3A_{2g} \rightarrow {}^3T_{1g}$ also shares the same trends (Fig. S8).

The extreme dichotomy between the delocalized photocarriers and localized excitons allows the time-resolved terahertz (THz) spectroscopy to conveniently distinguish them. [49-52] The detail of THz setup is described in Supplementary Materials. The THz electric field transmitted through the sample is denoted as $E(t)$, where $t$ is the delay between the THz generation and the electro-optical sampling pulses. The pump-induced change in $E(t)$ was then collected, denoted as $\Delta E(t;\Delta t)$, where $\Delta t$ is delay between the pump and THz pulses. To achieve a greater signal-to-noise ratio, the THz kinetics, $\Delta E(0;\Delta t)/E(0)$, is monitored at $t$=0 [the position of the maximum field strength (inset of Fig. 3e)]. The fast decay of the THz kinetics (Fig. 3e) directly measures recombination or the conversion from photocarriers to localized excitons.[51,52] As the existence of sub-ps recombination channels has been ruled out by the TR results, the THz kinetics reflects the photocarrier localization, and the localization rate agrees with that extracted from the TR measurement, which further validates the TR spectral assignment.

To further explore the driving force of the localization, first-principles calculations were performed in a 160-atom (4×4×1) supercell of $NiPS_3$ with a doped electron or hole. As the theoretical description of excitonic polarons is much more complicated than charged polarons within a single-particle picture, here we just consider an excess charge carrier in the supercell in order to simplify the computation and qualitatively evaluate the polaronic effect from a thermodynamic viewpoint, and this simplification has been shown to be effective in previous work.[53] The calculation reveals that a doped electron (hole) intends to localize due to the underlying antiferromagnetic background with fixed atomic positions, and the stabilization energy gained from localization is computed as 20 meV (10 meV). The carrier localization leads to a spin-flip of the nickel ion hosting this charge, associated with modifications of the magnetic moment of the neighboring ions (Fig. S9). When the lattice is allowed to relax, the doped electron (hole) prefers to polarizing the surrounding lattice structure and collapsing into a polaron at a $NiS_6$ octahedron with extra distortion (Fig. S10), leading to an increase in stabilization energy. Thus, the localization of $X_{AFM}$ excitons in the antiferromagnetic $NiPS_3$ might be driven by the self-induced distortion in both lattice and magnetic structures.

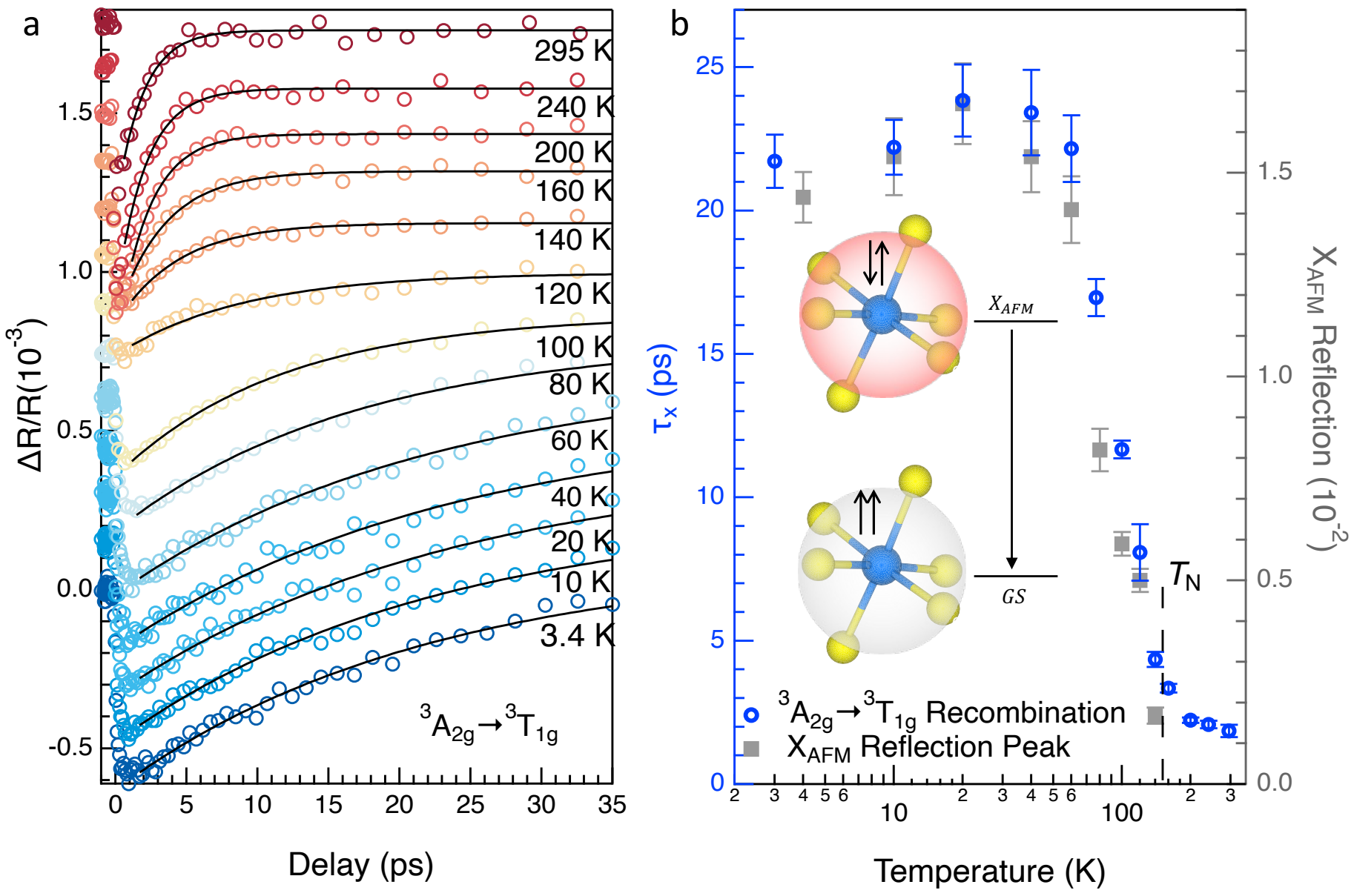

**Figure 4. Temperature dependence of exciton lifetime.** (a) The recombination kinetics of the localized excitons at different temperatures. (b) Temperature dependence of recombination characteristic time (blue circles). Also compared is the temperature-dependent $X_{AFM}$ reflectance (magnon sideband, grey squares) obtained from steady-state reflection spectra. Inset schematically illustrates the spin-state change involved in the exciton recombination.

**Correlation between exciton dynamics and the antiferromagnetic order.** To uncover the connection between the exciton dynamics and the magnetic order, we examine the exciton formation and recombination rates throughout the temperature range that covers both antiferromagnetic and paramagnetic phase zones. The TR spectra at various temperatures were displayed in Fig. S11. The photocarrier localization (or exciton formation) is measured by tracking the BGR decay, and its rate is constant throughout the antiferromagnetic and paramagnetic zones (Fig. S12). Because the $X_{AFM}$ bleach vanishes above $T_N$, the exciton recombination dynamics is probed at the resonance of $^3A_{2g} \rightarrow {}^3T_{1g}$, and the exciton lifetime ($\tau_X$) is determined from a single exponential fitting of recombination kinetics. In stark contrast with the localization, the exciton recombination is sensitive to temperature (Fig. 4a). The recombination is relatively slow and nearly constant at $T \ll T_N$ and then dramatically accelerates when the temperature approaches $T_N$ . As shown in Fig. 4b, $\tau_X$ is an order of magnitude greater in a typical antiferromagnetic phase than in a paramagnetic phase, and the temperature dependence of $\tau_X$ below $T_N$ resembles the temperature dependence of the long-range magnetic order represented by the $X_{AFM}$ reflectance.

Previous studies have been demonstrated that the exciton recombination in antiferromagnets can be explained quantitatively by the multimagnon emission model, and this recombination channel is governed by the spin-exchange interaction ($J$) and exciton energy gap ($\Delta$).[54-56] The temperature dependence of $\tau_X$ mainly arise from $\Delta(T)$.[55,57] Although $\Delta(T)$ for the exciton in antiferromagnetic $NiPS_3$ extracted from its optical bandgap shows a clear temperature dependence, the relative change of $\Delta(T)$ within the concerned temperature range is too small to account for the dramatic change of $\tau_X$ (Fig. S13). Thus, the multi-magnon emission channel is excluded as the dictating mechanism for the $X_{AFM}$ exciton recombination in $NiPS_3$. Nevertheless, as $X_{AFM}$ rooted in the underlying antiferromagnetic background corresponds to a triplet-singlet excitation,[4,14,30-32] the spin state is different between the exciton state and the ground state. Thus, a spin-flip is involved in $X_{AFM}$ recombination (inset, Fig. 4b), which should account for the suppressed recombination below $T_N$ due to the spin conservation rule. In the paramagnetic phase with disordered spin configurations, the localized excitons formed via on-site Coulomb interactions are no longer correlated with the spins, which are, in essence, multiples of correlated *d*-orbitals. In this scenario, the exciton recombination corresponds to spin-allowed transitions (e.g., ${}^3T_{1g} \rightarrow {}^3A_{2g}$ or ${}^3T_{2g} \rightarrow {}^3A_{2g}$), and the recombination bottleneck stemming from the spin-flip is removed. Thus, the connection between $\tau_X$ and the magnetic ordering can be qualitatively explained by the singlet character of $X_{AFM}$ inherited from the antiferromagnetic background. A quantitative description of the $X_{AFM}$ recombination rate requires a sophisticated theoretical model, which is unfortunately beyond the scope of this work.

In summary, this work deciphers the localized nature of $X_{AFM}$ and discloses the correlation between the $X_{AFM}$ lifetime and the magnetic order. We find that the magneto-exciton coupling determines not only the optical properties but also the exciton dynamics. According to the first-principles calculation, $X_{AFM}$ localization is mainly driven by deformations of the magnetic and lattice structures in the underlying antiferromagnetic order. Our finding offers potential opportunities to exploit the optical control for high-speed spin operation.

**Acknowledgements.** Y.Y. acknowledges Fundamental Research Funds for the Central Universities under Grant number 20720220011 and 20720240150, the National Key Research and Development Program of China (2022YFB3803304), and the National Natural Science Foundation of China under Grant Nos. 22175145. X.W. acknowledges the support of the National Key R&D Program of China (2022YFA1602704), the National Natural Science Foundation of China (62275225), the Fundamental Research Funds for the Central Universities under grant (20720220034), and the 111 Project (B16029). Z.L. acknowledges the Seed Fund from the Ershaghi Center for Energy Transition (E-CET) at the Viterbi School of Engineering, University of Southern California.